\PassOptionsToPackage{cmyk,table}{xcolor}
\documentclass[conference,flushend]{iaria}
\pdfoutput=1 

\usepackage[babel=true,english=american]{csquotes}
\usepackage[english]{babel} 

\usepackage[shortcuts]{extdash} 

\usepackage[alwaysadjust]{paralist}
\usepackage[caption=false,font=footnotesize]{subfig}

\usepackage{tikz}
\usetikzlibrary{shapes.geometric, arrows.meta, positioning, calc}

\usepackage{booktabs}
\usepackage{tabularx}
\usepackage{amsmath}
\usepackage[linesnumbered,plain, vlined]{algorithm2e}
\usepackage{graphicx}
\usepackage{soul}

\usepackage[
babel=true, 
expansion=alltext,
protrusion=alltext-nott, 
nopatch=eqnum, 
final 
]{microtype}

\usepackage{fontawesome} 
\usepackage{relsize}     
\usepackage{lipsum}      

\title{A Methodology for Investigating AI Patterns Prevalence in Software Repositories}

\author{
    \IEEEauthorblockN{Srinath Perera\textsuperscript{1}, Hasinthaka Piyumal\textsuperscript{1}, Frank Leymann\textsuperscript{2}, and Rania Khalaf\textsuperscript{1}}
    \IEEEauthorblockA{WSO2\textsuperscript{1}, University of Stuttgart\textsuperscript{2}}
    \IEEEauthorblockA{Santa Clara, CA, USA\textsuperscript{1}, WSO2, Stuttgart, Germany\textsuperscript{2}}
    \IEEEauthorblockA{e-mail:  \{srinath, hasinthaka, rania\}@wso2.com \textsuperscript{1}, frank.leymann@iaas.uni-stuttgart.de\textsuperscript{2}}
}

\begin{document}

\maketitle

\begin{abstract}
As Artificial Intelligence(AI)-based applications take off, a clear understanding of AI patterns can uplift the quality of AI applications. Many AI patterns have been proposed in the literature; however, their prevalence in real-life code has not yet been validated. Understanding the actual use of those patterns in practice can clarify our understanding both of the significance of these patterns and their utility. In this paper, we present a methodology to a) identify relevant patterns by mining the literature and then to b) validate their presence and prevalence in actual code repositories using active learning. To that end, we identify 14 AI pattern classes by mining 44 published AI pattern-related sources. Then we use an active learning approach to determine the prevalence of the most common pattern class across 100 GitHub open AI repositories. Using prevalence estimation, we propose bounds on the accuracy of the occurrences. The model achieves 56\% accuracy and 55\% recall in an 8-way classification task, significantly outperforming the 11\% random-chance baseline. Furthermore, the prevalence estimation offers usable bounds for analyzing pattern applications. This methodology provides a robust foundation to start understanding how AI patterns are used in practice, a field that currently lacks empirical data.
\end{abstract}


\section{Introduction}

Large Language Models (LLMs)~\cite{26} now provide general-purpose, high-quality Artificial Intelligence (AI) capabilities that require little to no user-provided training data. These models have unlocked many previously infeasible use cases. Several patterns and abstractions including Retrieval-Augmented Generation (RAG)~\cite{30}, ReAct~\cite{28}, and agent-based frameworks~\cite{29} now help developers build AI applications. Software Design patterns~\cite{27} document recurring problems and solution templates, thereby improving software quality by enabling, communicating, and educating best practices. 

As we will discuss in the related work section, many AI patterns have been proposed in the literature. Those patterns aim to capture the authors' observations and experiences. Validating those patterns by understanding their usage frequency in practice can clarify their relative importance, thereby improving the quality of AI-based applications. Furthermore, estimating usage frequency will help us verify candidate patterns using the rule of three (rule of X)~\cite{63}. 

We have identified 769 AI design pattern candidates proposed in the literature. Using word embeddings of those pattern descriptions and clustering, we have grouped them into 78 refined pattern candidates. Then, by manual inspection, we have categorized them into 14 pattern classes. 

To verify the prevalence of pattern classes in practice, we selected 100 open-source GitHub repositories that implement real-world AI applications. From those repositories, we have extracted 2442 code communities. (A code community is a group of tightly coupled methods (procedures) in the call graph.) 

Then, using an active learning-based approach~\cite{65}, we built a classifier to detect the most common patterns listed above. Active learning builds a model with the help of an (e.g., human) oracle, aiming to minimize the number of data annotations (i.e., oracle queries). We proceed by starting with an initial model and iterative labeling of a few data-points that exhibit higher uncertainty in their predictions under the current model.  Then, we rebuild the model with the new data and iterate. Consequently, we obtain an accuracy bound for how many times each of the above pattern classes will occur in the repositories. 

The paper makes the following contributions.

\begin{itemize}
    \item By combining methods active learning and human-in-the-loop annotations, we propose a robust methodology to detect patterns without pre-existing annotated datasets, thereby addressing a significant bottleneck in software engineering research: labeled data for niche or emerging patterns is rare.  
    \item We propose a new chunking method to generate code embeddings useful for pattern detection by applying the Louvain method~\cite{54} to call graphs to identify "code communities" that serve as chunks. 
    \item We propose applying prevalence estimation techniques (matrix inversion and Monte Carlo simulations) to estimate the true frequency of patterns with useful error bounds.
    \item We synthesized 769 pattern candidates into a refined taxonomy of 14 pattern classes. We provided one of the first empirical validations of AI pattern prevalence in real-life code by analyzing 100 open-source repositories.
\end{itemize}

The remainder of the paper is organized as follows. The section II discusses related work, and the following Section III describes the proposed methodology. Section IV describes the implementation details, followed by Section V, which describes the results. The section VI discusses the findings and lessons learned, and Section VII concludes the paper.  

\section{Related works}

This section discusses previous work on AI patterns and techniques for Mining Patterns from Code. 

\subsection{AI Patterns}
Researchers and practitioners have proposed many AI patterns. Among post-LLM patterns, Huang~\cite{31}  is a book on LLM design patterns. Gullí~\cite{48} and Liu et al.~\cite{47} discuss agent design patterns. Subramaniam ~\cite{52} and Jain~\cite{35} provide the practitioner's perspective. The AWS web page~\cite{34} provides a detailed description of best practices for implementing AI and ML applications on AWS. Also, Databricks documentation~\cite{45} lists a collection of Agent system design patterns.  Furthermore, Arslan et al.~\cite{32} and Singh et al.~\cite{49} provide detailed surveys of RAG patterns; articles~\cite{50} and~\cite{52} present practitioners' perspectives on the same pattern. 

Among pre-LLM patterns, Lakshmanan et al.~\cite{41} is a textbook on design patterns from the ML domain. Nalchigar~\cite{53} provides a detailed taxonomy of solution patterns for ML tasks such as clustering, classification, and anomaly detection. Washizaki~\cite{44} lists 15 design patterns (derived from a literature study) for building ML applications. Rodaz et al.~\cite{40} propose a mathematical formalism that sketches tasks around ML, but it is not digestible by developers. The Azure website~\cite{3} lists best practices for MLOps. 

You can find a comprehensive list of sources we used to mine patterns from~\cite{64}. These include the sources mentioned above, as well as additional practitioner content, such as articles, blogs, and documentation.

\begin{figure*}[ht!]
    \centering
    \includegraphics[width=0.8\textwidth]{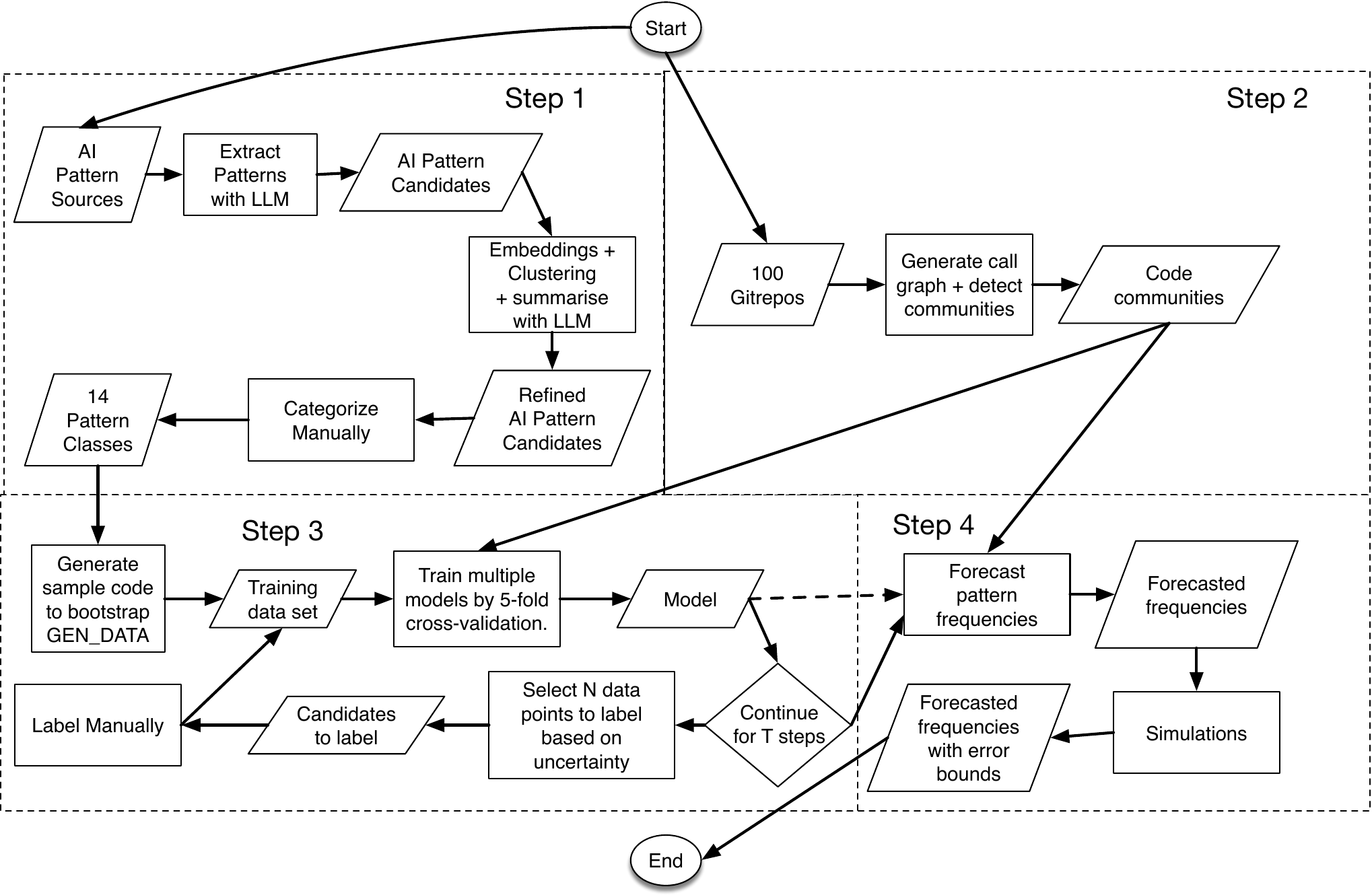}
    \caption{Proposed Methodology}
    \label{fig:PatternExtractionMethodology}
\end{figure*}

\subsection{Mining Pattern from Code}
When detecting patterns in code, we need to first find a code representation and a pattern representation, and then match them. Calculating code representations involves a trade-off between losing information and maintaining generalizability. Pattern representations closer to code are often suboptimal as they fail to generalize across multiple languages. On the other hand, manually crafted pattern representations require significant effort and scale poorly to large numbers of patterns. Many techniques have been proposed for finding design patterns from code. 

Search-based approaches use representations closer to code. Work such as Kramer et al.~\cite{20} and Dabain et al.~\cite{21} focused on detecting patterns using rules that describe them. These rules detect relationships between classes (e.g., inheritance hierarchies and method invocations). To reduce the overhead of pattern specification, Ghulam et al.~\cite{23} break GOF~\cite{27} patterns into 44 reusable atomic patterns, detect each using SQL, Regular Expressions, or AST parsing, and combine them to discover the patterns. Zdun et al.~\cite{57} turn microservice patterns into formal rules and provide a tool that automatically evaluates code in terms of compliance with these patterns. 

A logical extension of this method is to represent patterns as subgraphs and to search the code graph for them. Mayvan~\cite{10} is an example of this approach. On the other hand, Tsantalis~\cite{17} used similarity scores between the code graph and the pattern subgraph, enabling more flexible matching. GEML~\cite{7} evolves a set of human-readable if-then-else rules using a greedy algorithm based on inexact graph matching to detect a pattern.

More recent work often focuses on ML-based statistical methods, which can avoid challenges of specifying patterns using rules or templates. Instead of specifying the pattern, they learn a pattern-identifier from sample data. 

Among them, classification techniques such as those of Uchiyama et al.~\cite{18} and Dwivedi et al.~\cite{9} trained a classifier on software metrics such as Depth of Inheritance, Coupling, the number of methods, etc. Chihada et al.~\cite{24} use features extracted from an association graph to train the classifier. Zanoni et al.~\cite{19} use a combination of graph matching and ML techniques. 

The downsides of initial ML techniques include the need for feature engineering, loss of information in features, and limited availability of training data.

Najam et al.~\cite{6} build a representation of the source code, apply the word2vec algorithm to construct an embedding space, and then use a classification algorithm. Pandey et al~\cite{3} use code embeddings (RoBERTa) generated from code with a KNN algorithm to achieve an F1 score of 0.91. This approach is very flexible as it can be used with minimal effort to detect new patterns if labels for the data are available. 

Most of the above approaches work with benchmarks P-MART~\cite{15} and DPB~\cite{16}, which primarily focus on software engineering patterns (e.g., GOF). No suitable dataset is available for detecting AI patterns. 

Furthermore, a common limitation is that most benchmarks focus solely on GOF patterns. One notable exception is Fernández et al.~\cite{25}, which first builds quantum circuits and then uses state-machine-based pattern-detection methods. 

Our work is motivated by the results of Pandey et al.~\cite{3}, which demonstrated the viability of embedding-based approaches. By retaining most of the information in code, embeddings avoid the need for feature engineering. However, embeddings are highly sensitive to the chunking method (i.e., which code is grouped together for calculating embeddings). We extended the work of Pandey et al.~\cite{3}, which uses P-MART with clean code segments each including a single pattern, to detect patterns in real-life code repositories. Also, our work extends beyond well-known patterns (e.g., GOF) and limited annotated datasets by addressing the lack of training data via an active learning approach. 

\section{Methodology }

Figure~\ref{fig:PatternExtractionMethodology} depicts the methodology we used to estimate the occurrence of AI patterns in real-world AI applications as a flow chart. It has four steps. Each of the following subsections discusses four steps of the methodology in detail. 

\subsection{Step 1: Extracting pattern candidates from the literature} 
For mining AI patterns, we used the sources listed in the related work and practitioner articles, blogs, and documentation. You can find the detailed list from~\cite{64}. For each source, we extracted text, and for videos, we used the transcript. 

We query an LLM (using Prompts~\cite{64}) to extract pattern candidates. We receive resulting pattern candidates in the template specified in the prompt. 

To detect similarity between pattern candidates, we generated word embeddings for pattern description and clustered embeddings using the DBSCAN algorithm. We explored clustering algorithms DBSCAN, HDBSCAN, OPTICS, Bayesian Gaussian Mixtures, and K-Means. We picked DBSCAN by evaluating the even distribution of cluster sizes via silhouette score. We summarize each resulting cluster using an LLM with the prompt~\cite{64} and use them as refined pattern candidates. 

Then, by manually inspecting pattern candidates, we categorize them under 14 pattern classes using the following criteria. 

\begin{itemize}
    \item We started with pattern classes: RAG, "Advanced LLM Prompting," "Forecasting with Classical Models," and "Multi-Agent Architecture," which were identified in related work. 
    \item We tried to identify a pattern class for each refined-pattern-candidate that we identified by clustering. If we failed, we created a new pattern class. Following this approach, we added pattern classes "Evaluating LLM Results," "Multimodal Data Processing and Prompting," "Using Tools with LLMs," "LLM-based User Intent Extraction," "LLM Fine Tuning," "Training \& Alignment," "Enabling Reliability, Explainability, or Robustness," "LLM-based Planning, XoT, ReAct, or Reasoning," and "MLOps" using this approach. 
    \item We also added 2 more pattern classes: "Preprocessing Text and Numerical Data" and "Model Abstraction," while manually labeling data in step four. 

\end{itemize}

The classification of patterns and their refined candidates can be found in~\cite{64}.

\subsection{Step 2: Collecting Real-world Code Communities}
A code community is a group of tightly coupled methods (procedures) in the call graph. To extract code communities, we curated a dataset of 100 repositories hosted on GitHub. The selection process filtered for Python repositories that possess between 250 and 1,000 stars. We manually verified each candidate to ensure it was a valid AI-related project. The complete list of selected repositories is provided in~\cite{64}.

Manual inspection of 50 repositories with over 1,000 stars revealed that the majority were foundational frameworks, such as TensorFlow. Consequently, we targeted the 250–1,000 star range to intentionally isolate the `application layer' of AI software, to understand how typical developers use AI in practice.

Embeddings will only work if chunks include full patterns and are reasonably small. For each repository, we generated a procedure call graph and detected communities using the Louvain method~\cite{54}.  Since all functions within a single community are closely coupled, code within each community is likely to be related. Operating under the hypothesis that such code communities contain complete patterns, we treated all code in each code community as a single chunk and generated code embeddings for each community using \texttt{gemini-embedding-001}. We call this dataset $D_{unlabeled}$.
 
\subsection{Step 3: Build a model with Active learning} 
We built a model to detect known code pattern classes by first creating a bootstrap version, selecting data for manual labeling via active learning criteria, and then iteratively rebuilding the model using five-fold cross-validation~\cite{62}. This cross-validation approach allows us to fully utilize all labeled data points.

To bootstrap the model, we prompted the LLM to generate multiple samples for pattern classes identified in step 1, which we will call $D_{init}$. 

The algorithm in Figure~\ref{algo:active-learning} explains the active learning based approach. At each step (algorithm 2, line 5), we dropped classes with fewer than 5\% of the data and 20 data points. We do not know the true labels; hence, we cannot control the distribution of labeled data without significant effort. In the tradeoff between stable results vs. forecasting more classes, we chose stability. As steps progressed and more data became available, more classes will become available.

\newcommand{\myAlgoFont}{\small} 
\SetAlCapFnt{\myAlgoFont}       
\SetAlCapNameFnt{\myAlgoFont}   
\SetAlFnt{\myAlgoFont}          
\SetKwSty{myAlgoFont}           
\SetFuncSty{myAlgoFont}         
\SetCommentSty{myAlgoFont}      
\SetArgSty{myAlgoFont}          

\begin{algorithm}[ht]
\caption{Active Learning based Pattern Identifier}
\label{algo:active-learning}
\centering
\DontPrintSemicolon
\SetKwFunction{FBuild}{build\_pattern\_identifier}
\SetKwFunction{FTrain}{train\_hybrid\_model\_cv}
\SetKwFunction{FFindUncertain}{find\_top\_n\_by\_uncertainty}
\SetKwFunction{FLabel}{label\_manually}
\SetKwFunction{FMap}{hanlde\_rare\_classes}
\SetKwInOut{Input}{Input}\SetKwInOut{Output}{Output}

\Input{$D_{init}$ (Initial generated data), $D_{unlabeled}$ (Unlabeled embeddings), $T$ (Total iterations)}
\Output{$M_{final}$ (Trained pattern identifier model)}

\BlankLine
$V_{orig} \leftarrow \emptyset$ \tcp*{Initialize original verified pool}
$V \leftarrow D_{init}$ \tcp*{Initialize current working pool}
$M_{list} \leftarrow \text{None}$\;

\For{$i \leftarrow 1$ \KwTo $T$}{
    $V \gets \text{handle\_rare\_classes}(V)$\;
    
    \BlankLine
    \tcp{Step 1: Train model using current pool}
    $M_{list} \leftarrow \FTrain{V}$\;
    
    \BlankLine
    \tcp{Step 2: Active Learning - Identify uncertain samples}
    $S_{hl} \gets \text{find\_uncertain\_topN}(M_{list}, D_{unlabeled})$\;
    
    \BlankLine
    \tcp{Step 3: Human-in-the-loop labeling}
    $D_{new} \gets \text{label\_manually}(S_{hl})$\;

    \tcp{Step 4: Update pools}
    $V_{orig} \leftarrow V_{orig} \cup D_{new}$\;
    $D_{unlabeled} \leftarrow D_{unlabeled} - S_{hl}$\;
    $V \gets V_{orig}$\;
}
\Return $M_{list}$\;
\BlankLine
\end{algorithm}

In \texttt{find\_uncertain\_topN()}function, we select data points to label from $D_{unlabeled}$ by forecasting them using current models, normalizing class probabilities of different models, averaging them to create overall predicted class probabilities, and performing margin sampling. Margin is the difference between the highest and second-highest classes in the forecast, and we select data points to label based on the smallest margin.

\begin{algorithm}[ht]
\caption{Building a model with Cross-validation }
\label{algo:cross-validated-model}
\small
\DontPrintSemicolon
\tcp{Pseudocode for train\_hybrid\_model\_cv(..)}
\KwIn{Verified Data $\mathcal{V}$}
\KwOut{Models $M$, F1 scores of models $F$}
\BlankLine

$\mathcal{N} \leftarrow \{\text{"lr", "knn", "svc"}\}$ \;

$C_{folds} \gets \emptyset$ \tcp{fold configs}
$\mathcal{F} \gets \emptyset$ \;
$O_{folds} \gets \emptyset$\;

\For{$f \leftarrow \text{split\_to\_folds}(\mathcal{V})$}{

    $D_{train} \leftarrow \text{f.train}$\;
    $D_{val} \leftarrow \text{f.validation}$\;
    $D_{test} \leftarrow \text{f.test}$\;
    
    \BlankLine
    $M_{list}, F_1 \leftarrow \text{train}(\mathcal{N}, D_{train}, D_{val}, \emptyset)$\;
    
    \BlankLine
    \tcp{use algorithm in Figure 4 to forecast}
    $O_{folds} \gets O_{folds} \cup \text{forecast}(M_{list}, D_{test})$\; 
    $C_{folds} \leftarrow \text{get\_configs}(M_{list})$\;
    $\mathcal{F} \leftarrow \mathcal{F} \cup \{F_1\}$ \;
}

$ \text{print\_results}(O_{folds}, D_{test})$

\BlankLine
\tcp{Aggregate results and train final ensemble}
$M_{list}, \_ \leftarrow \text{train}(\mathcal{N}, V, \emptyset, C_{folds})$\;

\BlankLine
\Return $\{M_{list}, \mathcal{F}\}$\;
\BlankLine
\end{algorithm}

Algorithm in Figure~\ref{algo:cross-validated-model} presents the logic for building a cross-validated model. And algorithm~\ref{alg:forecast} describes how to perform the final forecast as a weighted average of probability distributions, with weights based on F1 scores across folds. 

\begin{algorithm}[t]
\caption{Final Forecast}
\label{alg:forecast}
\small
\DontPrintSemicolon
\tcp{Pseudocode for forecast(...)}
\KwIn{$x$ Input data set, Models $M$, F1 scores of models $F$}
\KwOut{$Y$ The final predicted class labels}
\BlankLine
\tcc{Definitions:}
$\mathcal{C}$: The set of all possible class labels\;
$f_i \in F$: The $F_1$ score used as the raw weight for the $i^{th}$ model\;
$P(y=c \mid m_i, x)$: The probability that model $m_i$ assigns to class $c$ for input $x$\;

\BlankLine
\tcp{Ensemble Forecast Calculation}
$Y = \underset{c \in \mathcal{C}}{\text{argmax}} \sum_{i=1}^{|M|} \left( \frac{f_i}{\sum_{j=1}^{n} f_j} \cdot P(y=c \mid m_i, x) \right)$ \;
\Return $Y$\;
\BlankLine
\end{algorithm}



\subsection{Step 4: Pattern Occurrence Estimation based on the results} 
Let $m$ denote the model trained in step 3 to detect pattern classes in a code community.
We utilize the normalized confusion matrix derived from the model's cross-validation, denoted as $C$. Given the set of code communities $D$ (constructed in step 2), we define $O_D$ as the observed distribution of pattern classes in $D$ as predicted by $m$. 

Assume $T_D$ is the estimated true distribution of class in D. 

\begin{equation}
\label{eq:linear_solver}
\mathbf{O}_{samp} = \mathbf{C} \cdot \mathbf{T}_D \implies \mathbf{T}_D = \mathbf{C}^{-1} \mathbf{O}_{samp}
\end{equation}

Given $O_D$, we can find $T_D$ by using a linear solver. 

When estimating $T_D$ using our model, we face two sources of uncertainty: model misclassifications and sampling errors from $O_D$. We minimize the first error using equation ~\ref{eq:linear_solver}, and the second by sampling from a multinomial distribution to estimate the true frequencies, then performing a Monte Carlo simulation to obtain error bounds. Algorithm 4 depicts our approach. 

\begin{algorithm}[ht]
\caption{Inversion-based Estimation via Monte Carlo Simulation}
\DontPrintSemicolon
\SetAlFnt{\small}
\SetCommentSty{small}

\SetKwFunction{FNormalize}{normalize}
\SetKwFunction{FSample}{sample\_multinomial}
\SetKwFunction{FSolve}{linear\_solve}
\SetKwFunction{FClip}{clip}
\SetKwFunction{FPercentile}{percentile}

\KwIn{$C$ (Confusion Matrix), $\mathbf{O}$ (Observed counts), $\alpha$ (significance level), $N$ (Iterations)}
\KwOut{$[L, U]$ (Confidence interval bounds)}

\BlankLine
$\mathcal{R} \leftarrow \emptyset$ \tcp*{Initialize results collection}
$C_{norm} \leftarrow \FNormalize{C}$ \tcp*{Normalize columns to sum to 1}
$n \leftarrow \sum \mathbf{O}$ \tcp*{Total sample count}
$\mathbf{p} \leftarrow \mathbf{O} / n$ \tcp*{Observed frequencies}

\BlankLine
\For{$i \leftarrow 1$ \KwTo $N$}{
    \tcp{Sample from multinomial to simulate observation noise}
    $\mathbf{O}_{samp} \leftarrow \text{sample\_multinomial}(n, \mathbf{p})$\;
    
    \BlankLine
    \tcp{Solve linear system using Equation 1}
    $\mathbf{E} \leftarrow \text{solve}(C_{norm}, \mathbf{O}_{samp})$\;
    
    \BlankLine
    $\mathcal{R} \leftarrow \mathcal{R} \cup \{ \text{clip}(\mathbf{E})\}$ \tcp*{Store clipped estimates}
}
\BlankLine
$L \leftarrow \text{percentile}(\mathcal{R}, \alpha/2)$\;
$U \leftarrow \text{percentile}(\mathcal{R}, 100 - \alpha/2)$\;

\Return $[L, U]$\;
\BlankLine
\end{algorithm}

\section{Implementation Details}

At every step, we use Gemini 3 Flash as the LLM. All experiments are done in an Intel Core™ i7-10510U × 8 processor with 16.0 GiB running Ubuntu 24.04.3 LTS

We already discussed the choice of clustering algorithm used in step 1 in the methodology. Furthermore, we tried dimension reduction with PCA and UMAP. We received the best results with DBSCAN with UMAP. 

In step 3, for code embedding, we experiment with embedding methods Jina Embeddings v2 (Code), CodeSage Large v2, CodeRankEmbed, RoBERTa, CodeBERT, Voyage Code-2, Gemini Text Embedding 004, and selected Gemini Text Embedding 00,4, which provided the best classification F1-score while building a bootstrapping model in active learning. 

Unlike step 1, in step 3, using dimensionality reduction with the classifier model reduced the F1-score; Hence, we did not use any dimensionality reduction. 

While building the model for step 3, we tried Logistic Regression (LR), Neural Networks, Support Vector Classifier (SVC), Naive Bayes, Random Forest, XGBoost, Gradient Boosting, K-Nearest Neighbour (KNN), and Decision Tree. We used LR, SVC, and KNN, which provided the best individual F1 scores in the final averaging ensemble. 

While building the model in step 3, we achieved about a 4\% improvement in F1 score using the averaging ensemble method in code listing 3. Building a linear regression stacked model using outputs from a neural network and a linear regression model also yielded good results. Still, we chose the averaging ensemble method because the stacked model is complex and highly susceptible to overfitting.

In step 3, we used the following workflow for manually labeling code communities. Given a code community (cc) and a predicted pattern class pc, we use the LLM to generate a code description for cc, ask the LLM to judge whether the forecasted pattern class pc is correct, and describe its decision. We then manually verified that the LLM’s judgment was correct. If the prediction (pc) is wrong, we read the description of cc and propose an alternative pattern class.~\cite{64} lists Prompts used for this step. 

During active learning, in the third iteration, we tried building the model both with and without LLM-generated sample data, and it performed better with only verified data. So we only used verified data from the third iteration onward. 

In step 4, for the simulation, we assume the confusion matrix of the model developed in step 3 is representative. As discussed in the methodology, we model the sampling error as independent categorical draws from a multinomial distribution, with the observed category counts as the basis for the distribution. We used 20000 as the iteration count, increasing it until repeated tests did not change the outputs. 

\section{Results}

Manually verified data had pattern frequencies as follows: Forecasting with Classical Models (64), None (64), LLM-based Multimodal Generative Prompting (52), Preprocessing Text and Numerical Data (51), Using Tools with LLMs (31), RAG (28), Agent Architecture (20), Model Abstraction(20), Evaluating LLM Results (18), MLOps (18), Advanced LLM Prompting (14), LLM-based Planning, XoT, ReAct, or Reasoning (12), Enabling Reliability, Explainability, or Robustness (8), LLM-based User Intent Extraction (8), and LLM Fine-Tuning, Training \& Alignment (4).


%

However, the rest of the results will only have 7 pattern classes because the model will automatically map any class less than 5\% and 20 instance to the None class. Table ~\ref{tab:active-learnning-perf} shows the results after the fourth iteration.

\begin{table*}[ht!]
\centering
\begin{tabular}{llrrrr}
\toprule
& Category & Precision & Recall & F1-Score & Verified Data\\
\midrule
C1 & Agent Architecture & 0.50 & 0.30 & 0.38 & 20 \\
C2 & Forecasting with Classical Models & 0.69 & 0.79 & 0.74 & 63\\
C3 & Multimodal Generative Prompting & 0.62 & 0.73 & 0.67 & 52 \\
C4 & Model Abstraction & 0.45 & 0.50 & 0.48 & 20 \\
C5 & Preprocessing Text and Numerical Data & 0.53 & 0.50 & 0.52 & 50 \\
C6 & RAG & 0.67 & 0.64 & 0.65 & 28 \\
C7 & Using Tools with LLMs & 0.42 & 0.35 & 0.39 & 31\\
C8 & None & 0.56 & 0.54 & 0.55 & 85\\
\midrule
& \textbf{Overall Values} & \textbf{0.56} & \textbf{0.55} & \textbf{0.55} \\
\bottomrule
\end{tabular}
\caption{Classification performance metrics across categories.}
\label{tab:active-learnning-perf}
\end{table*}

The verified data column shows the number of manually labeled data points available. As expected, classes with fewer training samples have lower F1 scores. 

\begin{figure}[ht!]
    \centering
    \includegraphics[width=\columnwidth]{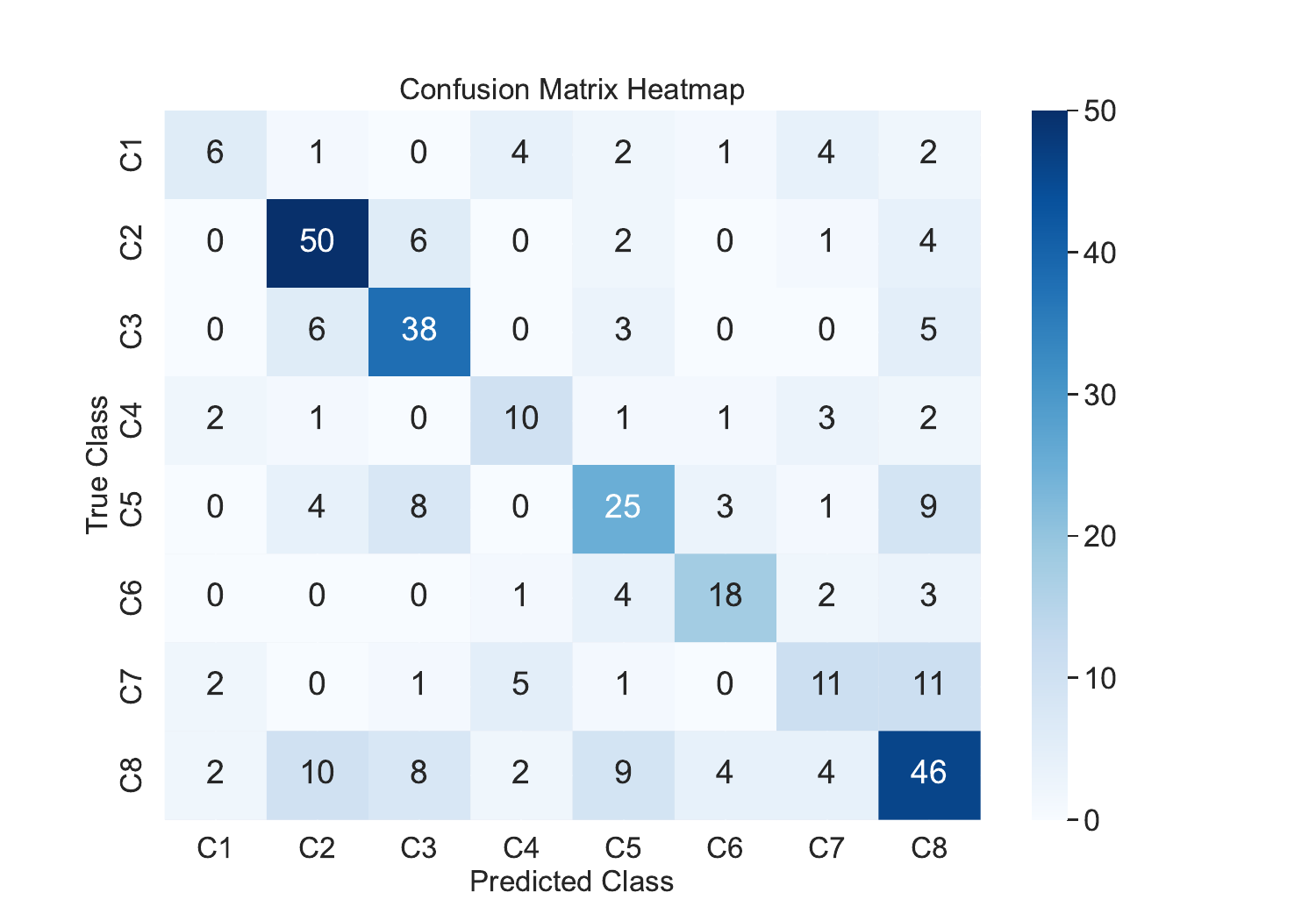}
    \caption{Weighted Vote Confusion Matrix}
    \label{tab:confusion_matrix}
\end{figure}

Table ~\ref{tab:confusion_matrix} shows the confusion matrix.

A strong main diagonal indicates the model's strength. As we discussed in the methodology, most classes are often misclassified as "None", and the "None" class often gets misclassified. Hence, it is a common source of error. Considering the often misclassified patterns, "Using tools with LLM" is often misclassified as "None", which could be because "tool use" is not very prominent in code and lacks a strong signal. Furthermore, classical models and preprocessing both get confused by "None" in both directions, which may be because these patterns include a wide variety of techniques that are harder to generalize with a small set of samples.

\begin{figure}[ht!]
    \centering
    \includegraphics[width=\columnwidth]{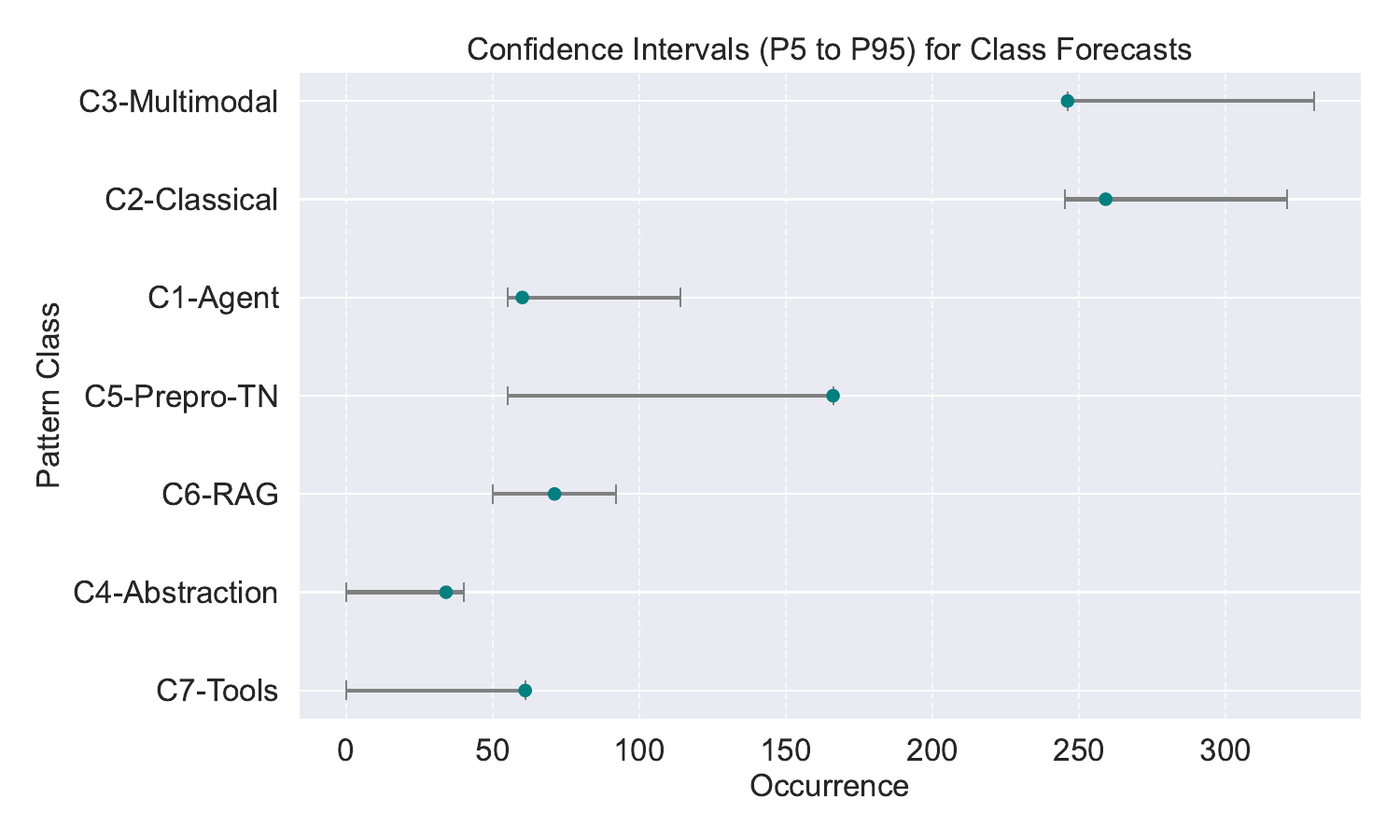}
    \caption{Estimated Pattern Class Prevalence with 95\% confidence intervals}
    \label{fig:prevalence-estimation}
\end{figure}

Figure~\ref{fig:prevalence-estimation} depicts our results from the prevalence estimation described in step 4. 

There are several zeros in bounds for the last two classes, which could occur when a class is affected by high false-positive noise from other classes or when it is often misclassified as other classes. We conducted a sensitivity test for both cases and found that the probable cause is that "Tool Use for LLMs" is often misclassified as "None" as shown in the confusion matrix in Figure~\ref{tab:confusion_matrix}. 

Smaller training samples likely cause wider confidence intervals for the last two classes. When we exclude those, the confidence interval variation is 45\%, which provides usable estimates for our problem. Furthermore, confidence levels indicate when estimates are unreliable, and users can address that by running more iterations of the algorithm. Since the algorithm selects labeling data based on uncertainty, there is a good chance that underrepresented classes will be labeled in subsequent iterations. 

Where are the prevalence results at the current accuracy level useful? First, it can confirm the wider availability of 5/7 pattern classes. Furthermore, it can give us relative availability of different pattern classes. For example, even using given confidence levels, we can argue that pattern classes C2 and C3 are much more common than C1, C5, and C6. Such understanding can help us focus our attention on teaching these patterns. However, the current accuracy level is not enough to identify a specific pattern given a code segment, which is a future direction we plan to explore.

The following lists concrete examples organized by pattern class, found based on the model's forecast for extracted communities.
\begin{itemize}
    \item Agent Architecture - multi-agent debate, multi-agent orchestration, communication through an agent bus, coordination through group chat.
    \item LLM-based Multimodal Generative Prompting - Predicting object centers,  dimensions, and local offsets, inferring an object's 3D position, bounding box prediction, motion prediction, sequential detections of an object from Lidar data, resizing and padding 3D tensors, generating synthetic training data.
    \item Model Abstraction - abstracting multiple AI backends or local/ cloud producers, routing, adopting Prompts via template, and managing API keys
    \item Preprocessing Text and Numerical Data - anonymizes, binning, type conversions, data dictionary, extracting metadata, text normalization, categorical feature encoding
    \item Retrieval Augmented Generation(RAG) - search with vector database, using custom knowledge bases, caching, and retriever component
\end{itemize}

\section{Discussion and Lessons Learned}
 
One code community can house multiple patterns. However, to keep manual data labeling simple, we have only used a single label. However, our output provides a probability distribution across all classes, which can be used to detect multiple patterns. 

Contrasting our results with those of Pandey et al.~\cite{3}, which achieved an F1-score of 0.91, it is worth noting that we also tried the same algorithms and embeddings. The likely difference is the curated nature of PMART vs. the complexity of real-life code we used, which often includes additional logic. Furthermore, as we discussed in the results section, users of our methodology can assess the reliability of estimates using error bounds. If they require tighter error bounds, they can likely achieve them by running more iterations of the step three active learning algorithm, which likely yields more training data for those classes and thereby improves the error bounds. Given the limited understanding of design pattern frequency in the real world, even the bounds we reported after four iterations remain usable for 5 out of the 7 classes and represent a significant step forward. 

In step 3, we folded classes that had fewer than 5\% or 20 samples into the None class, which is a practical choice. Some patterns are common, while others are rare, and naturally, there are many potential patterns. To make our approach stable, we had to draw the line somewhere. When the number of samples in a class is small, when combined with 5-fold cross-validation, the F1-score changed significantly (>10\%) due to each misclassified sample, thus becoming unreliable~\cite{68}.

However, we acknowledge that this practical choice reduces the granularity of results, as artificial non-class can interfere with other classes. Because the boundary between what is considered a pattern and what is not is fluid (e.g., subjective in how abstract we want the problem and solution to be), we believe our choice is still useful. If needed, practitioners can often get more granularity by labeling more data points.

We extend our work beyond basic software patterns (e.g., GOF) and limited annotated datasets by addressing the lack of training data via an active learning approach. The methodology we used is highly independent of AI patterns, and the same approach is likely to work well with other types of patterns. The ability to detect patterns by building a classifier on embeddings has been demonstrated by Pandey et al.~\cite{3} using the P-MART data set, and by this paper on real-world AI projects. It is likely that other domains also contain sufficient information in their embeddings. 

Detecting patterns via embeddings is highly sensitive to the choice of chunking method. Initially, we generated embeddings per file in the repository, resulting in poor classification performance in step 3. We needed a chunking method that would keep all relevant code about a pattern in a single chunk. Given that goal, typical chunking approaches like fixed size, pooling, by function, or by class would not work. AST-like methods (e.g., CAS~\cite{[66}) also focus on code syntax structure not the dependacies. In contrast, a call graph naturally captures dependencies similar to code graph-based approaches used in related work(e.g., Mayvan~\cite{10}, Tsantalis~\cite{17} ). It is much more likely that code from the same pattern, which is more closely related, is more connected in the call graph. This is also a natural extension of Codegrag~\cite{67}, which has used a similar idea for code retrieval. The performance improved significantly after adopting the call graph-based communities as chunks.

We tried to detect patterns by clustering the code communities collected in step 2 and then interpreting and summarizing the resulting clusters. This approach did not work well. However, it may be possible to fine-tune the embeddings using techniques such as contrastive learning improve the separation between clusters, which is also a future work we plan to explore. 

Methodology uses LLM in three ways, and each can introduce bias. First, pattern candidates were extracted using the LLM and manually curated to identify pattern classes. Manual curation should help reduce bias, but will not address bias shared between LLMs and humans. Second, although we bootstrapped with LLM-generated data in the first round, after 3rd round, we dropped LLM-generated data because only using labeled data gave better results. Third, LLM-assisted manual labeling can also introduce bias. Exploring the use of LLM as a judge technique to reduce such bias is a useful direction for future research.

\section{Conclusion and Future Work}
In this paper, we aim to enable and shift the exploration of AI design patterns toward empirical validation using software repositories. To that end, we address three challenges. Embedding-based pattern detection is highly susceptible to chunking methods. We proposed a new chunking method based on call graph-based community detection. The performance of the resulting model validates our method. Second, AI has limited annotated pattern datasets, which we address through an active learning-based approach. Third, we address both sampling
and classification errors while estimating pattern prevalence by using a matrix-inversion-based technique and Monte Carlo simulations. 

The model achieves 56\% accuracy and 55\% recall in an 8-way classification task, which is 5 times higher than the 11\% random-chance baseline. Prevalence estimation provided useful error bounds for 5 of 7 classes. In other cases, users can detect unreliable estimates through error bounds and discard them, or obtain better estimates by running more iterations in step three. 

We synthesized 769 pattern candidates into a refined taxonomy of 14 pattern classes. Most patterns show more than three examples required to pass the rule of three. Frequency estimates for those classes confirm our understanding in some cases (e.g., Preprocessing Text and Numerical Data and Classical Models) and open new avenues of inquiry in others (Multimodal prompting and RAG). 

This work opens several avenues for future research. First, applying the techniques in other domains to explore the universality of the approach (e.g., applying the same techniques for microservices, quantum computing), second, studying the relationships between possible bounds vs. the number of training samples for each class, and third, exploring techniques, such as SLM fine-tuning and constructive learning, to improve model performance. All information required to recreate the above work, including code, labeled data points from step 4, and code, is available in~\cite{64}. 

\printbibliography

\end{document}